\documentclass[11pt]{article}
\pdfoutput=1
\usepackage{amsmath}
\usepackage{amssymb}
\usepackage{fancyhdr}
\usepackage{bbold}
\usepackage{graphicx}
\usepackage{appendix}
\usepackage[sort&compress,square,comma,numbers]{natbib}

\usepackage[
	colorlinks=true,
	citecolor=black,
	linkcolor=black,
	urlcolor=blue,
	hypertexnames=false]{hyperref}
	
\newcommand{\arXiv}[2]{\href{http://arxiv.org/pdf/#1}{{\tt #2/#1}}}
\newcommand{\arXivold}[1]{\href{http://arxiv.org/pdf/#1}{{\tt #1}}}

\let\oldsqrt\sqrt

\newcommand{\beq}{\begin{eqnarray}}
\newcommand{\eeq}{\end{eqnarray}}

\def\sqrt{\mathpalette\DHLhksqrt}
\def\DHLhksqrt#1#2{%
\setbox0=\hbox{$#1\oldsqrt{#2\,}$}\dimen0=\ht0
\advance\dimen0-0.2\ht0
\setbox2=\hbox{\vrule height\ht0 depth -\dimen0}%
{\box0\lower0.4pt\box2}}
\allowdisplaybreaks[1]

\renewcommand{\part}[1]{\vspace{.10in}{\bf (#1)}}
\newcommand{\Kahler}{K\"ah\-ler~}

\begin{document}

\begin{titlepage}

\vskip.5cm \begin{center} {\huge \bf A Light Stop with a Heavy Gluino: \
\vskip 8pt
Enlarging the Stop Gap}\\ \vskip
8pt

\vskip.1cm \end{center} \vskip0.2cm

\begin{center} {\bf {Kevin F. Cleary} {\rm
and} {John Terning}} \end{center} \vskip 8pt

\begin{center} {\it Department of Physics, University of California, Davis, CA
95616} \\

\vspace*{0.3cm} {\tt cleary@ms.physics.ucdavis.edu, terning@physics.ucdavis.edu} \end{center}

\vglue 0.3truecm

\begin{abstract}
\vskip 3pt
\noindent 
It is widely thought that increasing bounds on the gluino mass, which feeds down to the stop mass through renormalization group running,  are making a light stop increasingly unlikely. Here we present a counter-example. We examine the case of the Minimal Composite Supersymmetric Standard Model which has a light composite stop. The large anomalous dimension of the stop from strong dynamics pushes the stop mass toward a quasi-fixed point in the infrared, which is smaller than standard estimates by a factor of a large logarithm. The gluino can be about three times heavier than the stop, which is comparable to hierarchy achieved with supersoft Dirac gluino masses. Thus, in this class of models, a heavy gluino is not necessarily indicative of a heavy stop. 
\end{abstract}

\end{titlepage}

\section{Introduction}
With the  increasing the bounds on the gluino mass \cite{Chatrchyan:2011zy,Aad:2011ib} natural supersymmetry (SUSY), at least in its most common realizations, appears to be in trouble. This is because the mass of the squarks are sensitive, through loop effects, to the mass of the gluino \cite{ArkaniHamed:1997ab}. A heavy gluino usually implies a heavy stop,  but a heavy stop implies a fine-tuning in order to get the correct Higgs mass. In this paper we will examine a class of models in which the UV sensitivity of the squark masses to the gaugino masses can be suppressed via  strong dynamics. 
 
In the Minimal Composite Supersymmetric Standard Model  (MCSSM) \cite{Csaki:2011xn,Csaki:2012fh} a naturally light stop can arise as part of a dual meson in the low-energy effective theory. In these models the light states needed to make SUSY natural are composites, which are kept naturally light, whereas the other superpartners, for instance first and second generation squarks, are elementary and can have much larger masses. This is a 4D realization of the type of sequestering that happens in SUSY Randall-Sundrum models \cite{Gherghetta:2011wc} where the SUSY breaking is on the UV brane with the elementary fields while the composites are localized on the IR brane.

In the renormalization group (RG) running of  the stop mass there is a term proportional  to the gluino mass, which tends to increase the stop mass as one evolves toward the infrared (IR). However, there is an additional effect, with the opposite sign, coming from the anomalous dimension of the stop.  In the region of strong coupling (roughly 2-10 TeV) the underlying strong dynamics means that the anomalous dimension of composite fields is ${\mathcal O}(1)$. These competing terms generically push the stop mass towards an IR quasi-fixed point \cite{Hill:1980sq}. The value of the quasi-fixed point mass is smaller than the usual estimate of the stop mass by a (large) logarithm.  This allows for a significant hierarchy between the stop and gluino masses. 

This paper is organized as follows. First, in section 2 we review the formalism of analytic continuation in superspace \cite{Cheng:1998xg,ArkaniHamed:1998kj,ArkaniHamed:1998wc,Luty:1999qc}  to compute the soft masses of the composites.  In section 3 we briefly review the MCSSM  and examine the RG running of the stop mass with both large anomalous dimension and perturbative gluino effects, and discuss how they can result in a hierarchy between stop and gluino masses. 

\section{Stop Mass in Composite Scenarios}

The masses of the elementary and composite fields are sensitive to the scale of SUSY breaking, $m_{UV}$, in the UV description. For the elementary fields this takes the usual RG solution form:
\beq m_{elem}^2\left(\mu\right) = m_{UV}^2 \left(\frac{\mu}{\Lambda}\right)^{\mathcal{O}\left(\alpha\right)}~,
\eeq
where the power depends on the perturbative anomalous dimension. For composite fields there are two types of contributions:
\beq m_{comp}^2\left(\mu\right) = m_{IR}^2+m_{UV}^2 \left(\frac{\mu}{\Lambda}\right)^{\gamma}~.
\eeq
The equation above is schematic, and in principle should involve the exponential of the integral of the anomalous dimension. The first term represents the fixed point of an RG flow for $\gamma>0$. Furthermore, the first term can be shown to vanish at the edge of the conformal window, a result we will reproduce below. This is relevant because the MCSSM is exactly at this point. The second term, which depends on a $\mathcal{O}\left(1\right)$ positive anomalous dimension, as we expect when the Seiberg dual has an IR fixed point, will drive this mass to zero rapidly as we approach the IR.

We next want to include soft SUSY breaking in the dual theory via the method of analytic continuation in superspace \cite{ArkaniHamed:1998kj,ArkaniHamed:1998wc,Luty:1999qc} in the presence of parametrically small SUSY breaking mass terms. To do this we introduce non-vanishing $F$ and $D$-term SUSY breaking spurions, and compute soft masses in this SUSY breaking background.
We first briefly review this formalism in SQCD, closely following the discussion in \cite{Csaki:2012fh}. The action for SQCD in the UV is given by:
\begin{equation} \mathcal{L} = \int d^4 \theta \, Q^\dagger \mathcal{Z}e^V Q + \int d^2\theta \,U \,W^\alpha W_\alpha~.
\end{equation}
We can introduce SUSY breaking spurions for scalar and gaugino masses by including higher components in superspace:
\beq  
\label{Zspurion}
\mathcal{Z} &=& Z\left(1-\theta^2\bar{\theta}^2 m_{UV}^2\right)~,\\
U &=& \frac{1}{2g^2} -i\frac{\theta_{YM}}{16\pi^2} + \theta^2 \frac{m_{\lambda}}{g^2}~.
\eeq
The holomorphic scale of the gauge theory is related to $U$ by
\begin{equation} \Lambda_h = \mu \,e^{-\frac{16\pi^2 }{b}U\left(\mu\right)}~.\end{equation}
Next we consider the action of anomalous rescalings on this theory. For the action to be invariant under transformations $Q\rightarrow e^AQ$ the wavefunction renormalization factor must have a (spurious) scaling like a real vector superfield:
\begin{equation} \mathcal{Z} \rightarrow e^{-\left(A+A^\dagger\right)}\mathcal{Z}~. \end{equation}
The holomorphic scale must also have a (spurious)  transformation due to the axial anomaly:
\begin{equation} \Lambda_h \rightarrow e^{\frac{2F}{b}A} \Lambda_h~. \end{equation}
It turns out to be convenient to introduce an additional redundant scale which is axially invariant:  
\begin{equation} \Lambda^2 =\Lambda_h^\dagger \mathcal{Z}^{\frac{2F}{b}} \Lambda_h~. \end{equation}

Mesons in the dual description are identified with quark bilinears in the electric theory. This implies that the mesons in the IR theory transform under axial transformations as $M\rightarrow e^{2A} M$. Imposing SUSY and axial invariance of the low energy action, one sees that the leading \Kahler term for the meson is given by:
\begin{equation} 
\mathcal{L}_K = \int d^4\theta \,\frac{M^\dagger \mathcal{Z}^2 M}{\Lambda^2}~.
\end{equation}
Taylor expanding in superspace this gives the $\mu \rightarrow 0$ value of the meson mass. We find the mass of the meson given in terms of $m_{UV}$ is given by:
\begin{equation} m_{IR}^2 = 2\left(\frac{3N-2F}{3N-F}\right) m_{UV}^2 
\label{simplemass}
\end{equation}
We see at the bottom edge of the conformal window ($F=3N/2$) there is no soft mass associated with the meson, a similar calculation produces analogous results for the dual quarks.

One can include higher dimension operators in the \Kahler potential, these lead to terms that give additional contributions to $m_{IR}$, but they are suppressed by additional powers of $\Lambda$, so for $m_{UV}\ll \Lambda_h , \Lambda$, these additional contributions are highly suppressed and we expect the (composite) dual mesons and dual quarks to have very small soft masses compared to elementary particles which have soft masses $\sim m_{UV}$ (possibly suppressed by loop factors depending on how the SUSY breaking is mediated).

\section{The Stop Mass in the MCSSM}

We will now briefly review the composite SUSY model discussed originally developed in \cite{Csaki:2011xn}, and further discussed in \cite{Csaki:2012fh}. The Minimal Composite Supersymmetric Standard Model (MCSSM) features composites Higgs and tops, with partially composite $Z$ and $W$s. This model sites at the edge of the conformal window so the soft masses of the dual quarks and meson are zero at leading order, as we saw in the previous section . The matter content of the electric description of this theory is as follows:
\beq
\begin{tabular}{ c|c|cccc } 

 & $SU\left(4\right)$  & $SU\left(6\right)_1$ & $SU\left(6\right)_2$ & $U\left(1\right)_V$ & $U\left(1\right)_R$ \\ 
  \hline
$ \mathcal{Q} $& $\Box$ & $\Box$ & 1 & 1 & $\frac{1}{3}$\\ 
$ \overline{\mathcal{Q}} $& $\overline{\Box}$ &1 & $\overline{\Box}$ & -1&$\frac{1}{3}$\\ 
\end{tabular}
\label{quarks}
\eeq

\begin{figure}[h]
    \centering
    \includegraphics[scale=0.6]{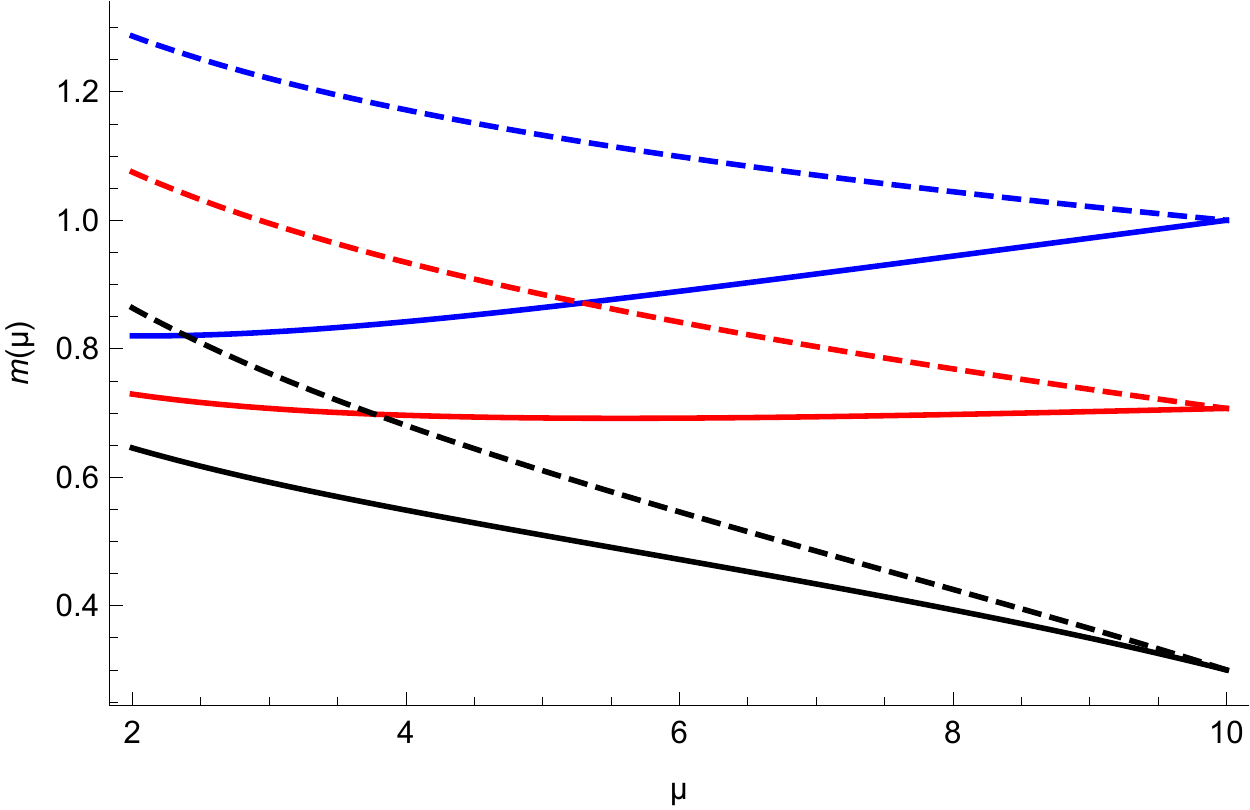}\includegraphics[scale=0.6]{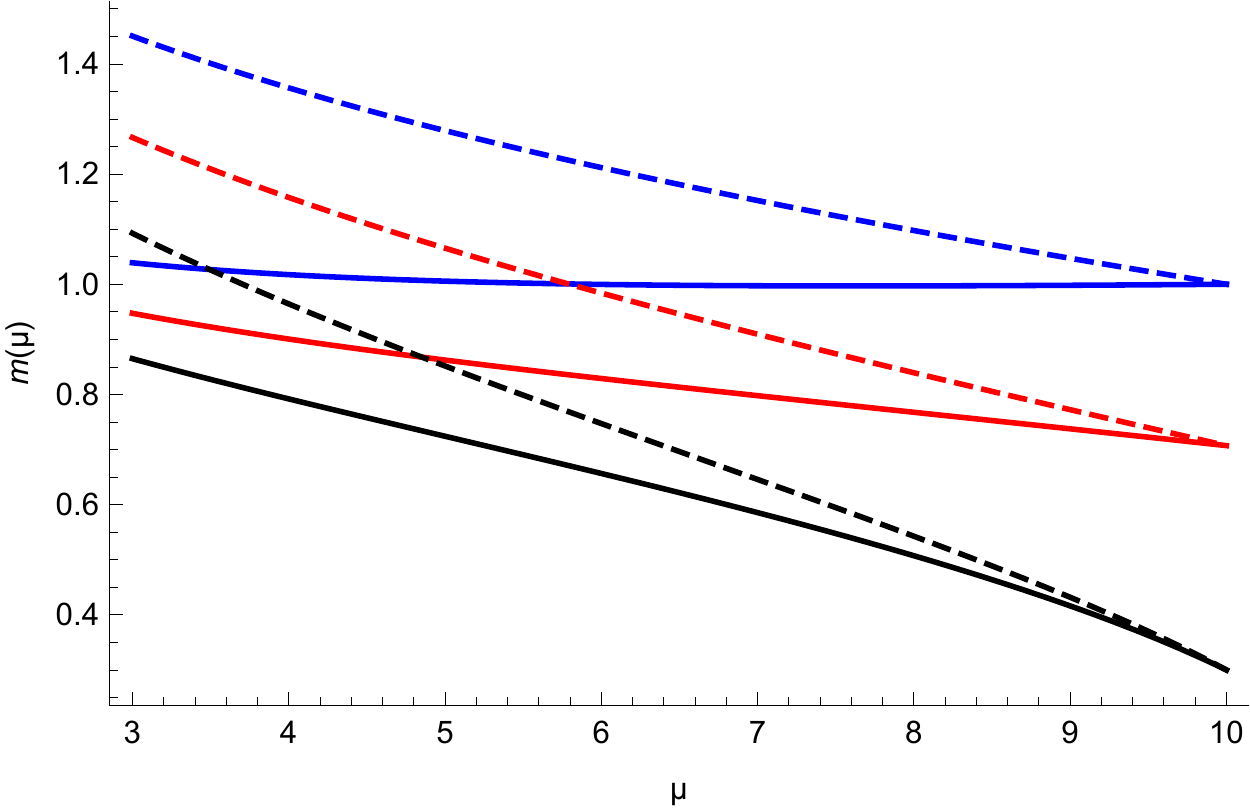}
    \caption{The stop mass, $m_{\tilde{t}}\left(\mu\right)$, in units of TeV, on the left with a 2 TeV gluino, on the right with a 3 TeV gluino. Dashed lines represent RG evolution ignoring the stop anomalous dimension, solid lines show RG evolution including the anomalous dimension.  Different cases correspond to different UV boundary conditions. The top pair of lines is for $m_{\tilde{t}}\left(10 \,{\rm TeV}\right) = 1 \rm{TeV} $; the middle pair is for $m_{\tilde{t}}\left(10 \,{\rm TeV}\right)= 700 \rm{GeV}$; and the bottom pair of lines is for $m_{\tilde{t}}\left(10 \,{\rm TeV}\right) = 300 \rm{GeV}$. }
    \label{fig:strong_running}
\end{figure}

The $SU\left(4\right)$ is the strong gauge group, and some of the flavor symmetries are weakly gauged to give an embedding of the SM. The magnetic dual of this theory is given by Seiberg duality \cite{Seiberg:1994pq}.
\beq
\begin{tabular}{ c|c|cccc } 
 & $SU\left(2\right)_{mag}$  & $SU\left(6\right)_1$ & $SU\left(6\right)_2$ & $U\left(1\right)_V$ & $U\left(1\right)_R$ \\ 
  \hline
$q$& $\Box$ & $\Box$ & 1 & 1 & $\frac{2}{3}$\\
$ \overline{q} $& $\overline{\Box}$ &1 & $\overline{\Box}$ & -1&$\frac{2}{3}$\\ 
$M$ & 1 & $\Box$ & $\overline{\Box}$ & 0 & $\frac{2}{3}$\\
\end{tabular}
\eeq
The dual theory has a superpotential, $W_{dyn} = y\, \overline{q} M q$. The SM is embedded in the global symmetries in the following way.
\begin{equation} SU\left(6\right)_1 \supset SU\left(3\right)_c \times SU\left(2\right)_{el}\times U\left(1\right)_Y~,\end{equation}
\begin{equation} SU\left(6\right)_2 \supset SU\left(3\right)_X \times SU\left(2\right)_{el}\times U\left(1\right)_Y~,\end{equation}
where the unfamiliar $SU\left(3\right)_X$ does not have to be gauged, and $SU\left(2\right)_{el}$ provides the elementary components of the $W^\pm$ and $Z$, after mixing with 
$SU\left(2\right)_{mag}$.
In this model the Higgses and the top are composite. The right-handed top is embedded in $q$, the up-type Higgs is embedded in ${\overline q}$, and the left-handed top is embedded in $M$. The leading IR soft masses,  as shown in Eq. (\ref{simplemass}), vanish, as we have emphasized repeatedly, since the theory is at the edge of the conformal window. 
If we assume that SUSY breaking is communicated by gauge mediation with the elementary strongly coupled quarks, $ \mathcal{Q} $ and $ \overline{\mathcal{Q}} $,   then we expect large, loop suppressed, soft masses for the elementary fields. The soft masses of the left or right stop will have RG equation contributions that depend on the stop anomalous dimension and the gluino mass.
\beq \mu \frac{d m_{\tilde{t}}^2}{d\mu} = \gamma \,m_{\tilde{t}}^2  -\frac{32}{3} \frac{\alpha_s}{4\pi}\,M_3^2~.
\label{RG}
\eeq
The first term is large at high energies, due to the underlying strong dynamics while the second term is the usual perturbative gluino contribution.
If we neglect the first term (as we would for an elementary field with a small anomalous dimension) then the  stop mass is very sensitive to the mass of the gluino. This is usually estimated to give a contribution of the size \cite{Csaki:2012fh} 
\beq
\Delta m_{\tilde{t}}^2 \sim \frac{32}{3} \frac{\alpha_s}{4\pi} M_3^2 \log \left(  \frac{\Lambda}{\rm  1 TeV} \right)~.
\label{estimate}
\eeq
Where $\Lambda$ is a UV scale, in our case it is around the duality scale.   However including the first term in (\ref{RG}) we see that (in the region of strong coupling) the stop mass is strongly suppressed by the first term. If $\gamma$ and $\alpha_s$ were constant there would in fact be a true fixed point. Since both run there is a quasi-fixed point, whose typical value is smaller than the estimate of (\ref{estimate}) by the factor of  the large logarithm.  We can estimate the quasi-fixed point value as
\beq
m_{\tilde{t}*}^2 \sim \frac{32}{3 \,{\bar \gamma}} \frac{\alpha_s}{4\pi} M_3^2 ~,
\label{newestimate}
\eeq
where ${\bar \gamma}$ is a suitable average of $\gamma$ over the running.  In Fig.~\ref{fig:strong_running} we plot representative solutions for  the running stop mass with a 2 or 3 TeV gluino. The dashed curves show the usual behavior for the RG running that includes only the gluino contribution. The solid curves include the strong dynamics as well, with the following ansatz for the anomalous dimension:
\beq
\gamma\left(\mu\right) = \gamma_{IR} + \frac{1}{2}\left(1-\gamma_{IR}\right)\left(1+\tanh\left(\frac{\mu - 1\rm{TeV}}{5 \rm{TeV}}\right)\right)~.
\label{anomalous}
\eeq
We have set $\gamma_{IR}$ to it's MSSM value $\sim 0.026$.

We see in all the examples that at low-energies the stop mass approaches a quasi-fixed point value that is much smaller than the gluino mass $\sim 2$ TeV. Typically, with this ansatz, we see a factor of three between the stop and gluino masses, similar to the factor of five hierarchy achieved in models with Dirac gluino masses \cite{Dirac}.



The suppression in the stop mass also leads to  a suppression in the  2-loop correction to the up-type Higgs mass due to the gluino loop is now
\beq
\Delta m_{H_u}^2 \sim -\frac{2 y_t^2 \alpha_s^2}{\pi^3}M_3^2 ~,
\label{higgs}
\eeq
which is smaller than the MSSM estimate by the square of a large logarithm. Thus there is a significant reduction in tuning.
Since the quartic Higgs coupling comes from the top Yukawa coupling in the MCSSM \cite{Csaki:2012fh}, the tuning in this class of models is roughly:
\beq
\frac{y^2 v^2}{2m_{H_u}^2}~,
\label{eq:finetuning}
\eeq
which (using (\ref{higgs})) is ${\mathcal O}(1)$ for a 2-3 TeV gluino, so there is no significant fine-tuning of the Higgs mass parameters in this type of model.

\section{Conclusions}

We have pointed out that the main impediment to natural SUSY theories---the sensitivity of the stop mass to the gluino mass combined with ever increasing bounds on the gluino mass---can be bypassed in composite SUSY scenarios. We have shown that in models like the MCSSM  we can generically have a significant hierarchy between the stop and gluino masses due to a quasi-fixed point in the RG flow of the stop soft masses. The splitting between the stop and gluino is comparable to what is seen in the very popular Dirac gaugino models. There is also no fine-tuning required to get the correct electroweak scale and Higgs. Thus, in general, it is still interesting to develop experimental search techniques \cite{Czakon:2014fka} to probe stop masses in the 200-800 GeV range that would allow a natural SUSY theory.

\section*{Acknowledgements}

We would like to thank C. Cs\'aki and A. Pomarol  for helpful discussions. J.T. is supported in part by the DOE under grant DE-SC-000999.

\end{document}